# A Security Plan for Smart Grid Systems Based On AGC4ISR

Ghazal Riahi

Faculty of Computer Science & IT, Payame Noor University,

Assaluyeh, Iran

Email: G.riahy@gmail.com

*Abstract* —**This paper is proposed a security plan for Smart Grid Systems based on AGC4ISR which is an architecture for Autonomic Grid Computing Systems. Smart Grid incorporates has many benefits of distributed computing and communications to deliver a real-time information and enable the near-instantaneous balance of supply and demand at the device level. AGC4ISR architecture is Organized by Autonomic Grid Computing and C4ISR (Command, Control, Communications, Computers and Intelligence, Surveillance, & Reconnaissance) Architecture. In this paper we will present a solution for as security plan which will be consider encryption, intrusion detection, key management and detail of cyber security in Smart Grids. In this paper we use the cryptography for the packet in the C4ISR and we use a key management for send and receive a packet in the smart grid because it is necessary for intelligent networks to keeping away from packet missing.**

## I.INTRODUCTION

The term "Smart Grid" [1] refers to a modernization of the electricity delivery system so it monitors protects and automatically optimizes the operation of its interconnected elements. It incorporates into the grid the benefits of distributed computing and communications to deliver real-time information and enable the near-instantaneous balance of supply and demand at the device level. This means that the Smart Grid will be characterized by a two-way flow of electricity and information to create an automated, widely distributed energy delivery network. Cyber security [9] to prevent unauthorized use of harm, exploitation [10], address and, if necessary, restoration of electronic information and communications systems and services (and the information contained therein) to ensure confidentiality [11], integrity [12] and availability.

The Smart Grid provides a reliable power supply and self-healing power systems, through the use of digital information, automated control, and autonomous systems. The Smart Grid is "green". It means that with reduce greenhouse gases and with supports renewable energy sources, and enables the replacement of gasoline-powered vehicles with plug-in electric vehicles. The Smart Grid [7, 8] continuously monitors itself to detect unsafe that could detract from its high reliability and safe operation. Higher cyber security is built in to all systems and operations including physical plant monitoring, cyber security, and privacy protection of all users and customers.

Additional risks to the grid include:

•     Increased use of digital information and controls technology to improve reliability, security, and efficiency of the electric grid [1].

•     Increasing the complexity of the grid could introduce vulnerabilities and increase exposure to potential attackers and unintentional errors [2];

• Dynamic optimization of grid operations and resources, with full cyber security [1];

• Increased number of entry points and paths for potential adversaries to exploit; and Potential for compromise of data confidentiality, including the breach of customer privacy [2].

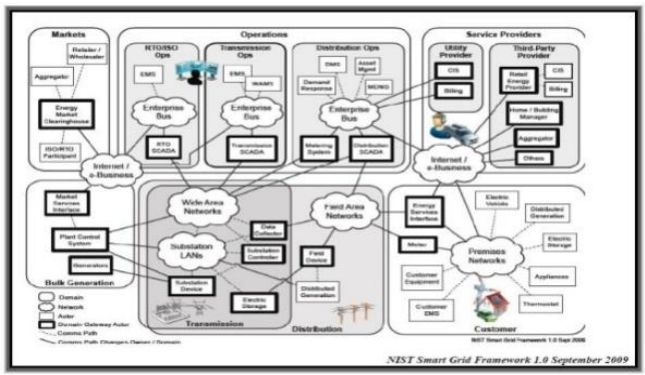

Figure 1 Smart Grid Framework[23]

Communication and collaboration services provide functionality to communicate and share information. Services include collection, processing and dissemination of information on the situation [23]. Operational data services and to evaluate the influence of other parties and also for the protection of state information about its status. Command and control services in order to support decision making and handling. Run and C4ISR systems connected and environmental factors are involved in the information flow control support obligation.

Control layer contains functions that support all of the features required by the above mentioned services and features such as security, mobility, and available used. Convergence layer ensures that it can connect to a unified manner based on Internet Protocol (IP) and a variety of fixed and wireless networks, belonging to the connection layer can be used is done.

Now we want explain some security strategy in smart grid and C4ISR. Also in this case we discussion about Encryption, Intrusion Detection, Key Management.

A. Encryption

The security of an encryption method should not depend on keeping the method secret (i.e. keeping the description of the algorithm secret), but should depend solely on keeping the key secret. To avoid attempts at smart grid activities, response metrics, capabilities, monitoring and detection techniques, voice and data encryption are critical at maintaining confidentiality of the operational information. Traditionally [3], this would be limited to the main communications channels, but in a C4ISR context these concerns should also address sensor data sources such as cameras, and biochemical detection devices. This requirement is known as the Kerckhoff Principle. These days [4], most encryption algorithms are actually made public and are described in standards and specifications. Decryption keys should be selected in such a way that it is not possible for an attacker to guess them.

Cryptanalyst point of view, the problem is the three variations of attacks. When he has a lot of encoded text and any text with just the encoded text has problems. Codes in the puzzle section of newspapers seems to have these kind of issues. When the cryptanalyst has some encoded text and the same text, the problem is known to be difficult text. Finally, when the cryptanalyst has the ability to encrypt text fragments and adequate facilities to prepare his defense, and he is now issue a text is selected. Newspaper cryptograms cryptanalyst can be broken down to detail allows you to ask questions such as "What is encryption ABCDEFGHIJKL?"

B. Intrusion Detection

There are several reasons why system-wide intrusion detection and reporting is important to DoD. The encoding strategies, mechanisms to enable C4ISR intrusion detection has become an imperative. Most of the monitoring system, traditional intrusion detection should be the integrated control system that ties into the infrastructure of C4ISR integration to provide alerts when a system is compromised when it has already been compromised prevent the compromised system without alerting the intruder, and helps coordinate the response to identify and arrest the intruder. First [24, 25], and obviously, when an intrusion is detected and reported, the considerable resources of the Department of Defense can be brought to bear to trace the intruder, to prevent or minimize damage, and to end the intrusion. Second, following neutralization of the intrusion, vulnerabilities in given systems can be remedied not only at the location of the intrusion, but throughout the DoD and other governmental locations where similar systems may be in operation. Finally, a widespread information warfare threat to national security as opposed to discrete localized challenges to individual operations and systems can only be identified if intrusions are detected and reported on a system-wide basis.

### C.  Key management

A public-key and symmetric key combined approach is proposed for simplicity and scalability of key management as well as other desirable properties. Secret and private keys from disclosure and modification should always be protected for the entire life cycle of a button. When creating keys, weak keys should be rejected. The transition between the two partners have a secret key for symmetric encryption via a secure channel. The keys are stored securely. If the key is no longer needed, it should be destroyed in a reliable manner. The symmetric key [17] scheme is based on the Needham-Schroeder authentication protocol, and the public key scheme is based on elliptic curve cryptography for high efficiency and strong security. The use of public keys also has a nice property that no static symmetric key is needed between data aggregators and collectors; this eliminates the possibility that symmetric keys could be compromised, and it also avoids the overhead of managing symmetric keys.

Now in this paper we want to use security platforms that used both of cryptography and key management. Because it is the important architecture [21, 22] and just one security plan it is not enough. so we have to use both of cryptography and key management in the same time.

## II.  SECURITY PLATFORMS

In this case we want discussion about Cyber security in C4ISR system and Smart Grid system.

### A.  Cyber Security in C4ISR Systems

Any computer network or information system is vulnerable to two types of attacks i.e. passive attacks and active attacks. The passive attacks can be traffic analysis and release of message contents. Passive attacks are generally very difficult to detect, however, encryption is a common technique to prevent passive attacks. Active attacks involve modification of data streams or the creation of false streams. These types of attacks fall into four categories i.e. masquerade, replay, modification of messages and denial of service (DoS)[15, 16]. Appropriate security mechanisms are designed and used to detect, prevent, or recover from a security attack. Various security mechanisms are encipherment, digital signatures, access control, data integrity, authentication exchange, traffic padding, routing control, notarization, trusted functionality, security label, event detection, auditing, recovery, etc.

Nowadays, the increasing reliance on information technology, computer and communication for military operations there. These factors have made the enemies of the hot IT infrastructure. C4ISR all about integration, interoperability and networking in many systems, and it is such a large system as we can make it a "system" is. Hence, more and more opportunities to attack opponents. These forces C4ISR system designers additional measures against enemy attacks, intelligent and determined to make the system secure against this type of attack.

There are four general vulnerabilities that can be experienced in C4ISR systems. Those are unauthorized access to data, clandestine alteration of data, identity fraud and denial of service (DoS)[16]. All of these vulnerabilities may have the catastrophic effects on national interests of a country. For example, unauthorized access to data on a C4I computer may result in inflicting severe damage to a nation by obtaining and using classified or unclassified sensitive information by an adversary. Similarly, military planning may be severely affected if clandestine alteration of data on a C4I computer is done by the enemy. The identity fraud may result in modification of situational awareness through insertion of unwanted/changed information, issuing of fake orders, etc. All these will affect the morale and effective working of the defense forces. By application of DoS on a C4I system, the time critical operational planning and completion of tasks may be affected.

Security in C4ISR systems is a two dimensional problem. Firstly, it means the physical security of the facilities where different components/parts of the larger system of systems are installed. No one can do this better than the military. However, the second part which is more demanding and challenging is the information security (INFOSEC)[5 , 6] which may be called as cyber security; as it is not clearly understood at all the levels.

The C4ISR systems require appropriate measures and requirements for the vulnerabilities discussed above for achieving confidentiality, integrity and availability of data and maintaining system configuration through elaborate security guidelines and accountability of personnel authorized to access the information sources. Following security services are required to cater the security requirements[16]:

• Authentication: ascertaining the user identity through various means such as passwords, fingerprints, digital certificates, etc.

• Access Control: Permission or authority to perform specified actions as authorized in policy guidelines.

• Data Confidentiality: protection of data against unauthorized disclosure.

• Data Integrity: The assurance that data received and is same as transmitted by authorized user.

• Non-repudiation: Providing protection against denial by participants once participated in communication.

• Availability: Ensuring availability of a capability or system at all times or whenever desired/required.

The world has become more unpredictable and more unstable, and new risks have emerged, in particular computer attacks. Cyber security of digital information worldwide network connection and speed up the software is released. Document automation and paperless exchange fundamentally changed economic environment, social and political life of the United Nations, they are more vulnerable.

### B. Cyber Security for Smart Grid System

In Smart grids cyber security we want check that what is the important problem in this case. So now we should explain that:

*Cyber Security Model.* Like for any other network's security, the three main objectives that cyber security focuses on is availability, integrity, and confidentiality, that is, availability of power with integrity of information and confidentiality of customer's information.

• *Availability.* The reason why we have smart grid is "availability". The basic goal of our network is to provide uninterrupted power supply to the users and to match user requirements.

• *Confidentiality.* The grid network is responsible for the protection of a user's information. If the data is not protected, ample information about the user can be revealed to the attacker.

• *Integrity.* The messages received from the user end should be authenticated. The network must ensure the information is not tampered. Also, the source of message should be authentic.

Security for the Smart Grid information and control networks must include requirements for: [2]

• security policies, procedures, and protocols to protect Smart Grid information and commands in transit or residing in devices and systems;

• authentication policies, procedures, and protocols; and

• Security policies, procedures, protocols, and controls to protect infrastructure components and the interconnected networks.

*Privacy and security: T*he latest, a strong need for privacy and security for smart grid there. Confidentiality of information relating to individuals, for example, information about the location of vehicles and equipment or the use of energy. Traditional access control mechanisms are useful to prevent unwanted access. However, many situations in which it is granted access to the original data, but data is further limited. It also rules that may apply to specific data is released. Means used to express constraints and technical limitations of such should be supported. So, for example, in the case of policies should be adapted to allow the WWW is a network aware privacy. In addition, the smart grid, including participants with very high security requirements. Participants can access the destructive effects of the disaster. Therefore, the communication architecture should be a strong sense of security that provides participant at risk, while still allowing access to / from the network remaining (open) grid.

## III.    A SECURITY PLATFORM FOR SAGC4ISR

In order to obtain cyber security, we must use different keys for data encryption and secure. In this section, we study the various aspects related to encryption and key management. We first through the limitations of encryption issues and solutions relate to the encryption and then go.

### A.    Constraints

In this part we wand explain some of the restriction in the AGC4ISR [18] and SAGC4ISR [19] which are introduced by Mehdi Bahrami et al.:

Computational constraints. Residential meters limitations when it comes to computing power and the ability to store cryptographic material. Future device requires a fundamental cryptographic functionality, including the ability to support symmetric encryption for authentication. Using low cost hardware with embedded encryption is necessary, but not sufficient to achieve high availability, integrity and confidentiality of the SAGC4ISR architecture.

Channel bandwidth. Communication that will take place in a SAGC4ISR architecture will take place over different channels with different bandwidth. AES encryption which produce the same number of output bits as input bits. It can compress a bit too much, because they are encrypted and random in nature. If you need to compress the data, we need to do before encryption. Another factor to be considered based cryptographic message authentication code (CMAC), which was added as a fixed overhead to a message is typically 64 bits or 96 bits. The overhead itself is remarkable that we are dealing with SMS, because they require large bandwidth channel.

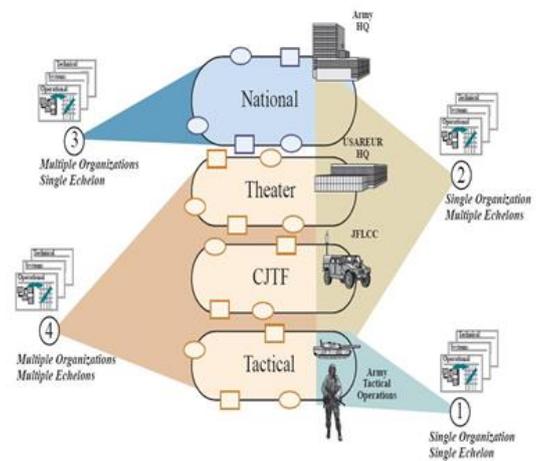

Figure 2 Four Dimensions of Architecture Integration [20]

There are four dimensions of architecture integration [20] that represent varying degrees of integration scope. Figure 2 illustrates these four dimensions in context with a global, hierarchical view of war fighter operations and support. Note that the need to integrate multiple architecture views and descriptions is certainly not limited to Joint or cross-organizational considerations. The Framework is intended to facilitate all four integration dimensions.

A first dimension involves a single organization and its operations within a single "echelon." This single organization is different from multi-level organization such as [8]. In addition to the obvious need to interrelate the three views (and associated products) of an Army tactical architecture, in this case there may be multiple architectures - at the same echelon that cover different functional areas or viewpoints that need to be interrelated, depending on the purpose and scope of the initiative. For example, the Army may be investigating more cost-effective means of providing logistics support to troops in the field. This may involve integrating the architecture views that reflect a war fighting perspective with the views reflecting a logistics-support perspective to assess tradeoffs between C4ISR and logistics investment options. We can use this dimension on autonomic grid by tactical architecture for manage as hierarchy in grid networks because some time we need mange autonomic node on the network.

A second dimension illustrated in figure 2 still involves a single organization (Army), but the integration scope expands vertically to include Army operations across multiple echelons. In this particular case, the organization may be examining opportunities to streamline its operations or investments from top to bottom.

A third integration dimension involves architecture initiatives that cross-cut multiple organizations (U.S. And/or multi-national) horizontally, within a single echelon. An example of this dimension is an architecture whose objective is to investigate opportunities for the various components of DoD to exploit or leverage National information infrastructure capabilities and in grid computing so.

A fourth dimension of integration involves multiple organizations and multiple echelons, where vertical and horizontal Joint relationships need to be articulated and examined. An example of this dimension is an architecture whose focus is on assessing the effectiveness of intelligence information support to the war fighter. This could involve examining tradeoffs between hierarchical support policies and practices, e.g., theater- based Joint Intelligence Center dissemination to lower-echelon users and direct dissemination from collectors to forces.

Connection. Many devices can connect to the key server, certificate authority certificates, online status, and no protocol server. Many SAGC4ISR architecture communications between devices connected to the Internet for longer than is typical.

B.    Issues of public cryptography

Entropy. Generation of cryptographic keys required to create an event that is a good source of entropy is unavailable for many devices.

Cipher suites. Cryptographic suite that is open about the need to achieve interoperability. Deciding on a block cipher mode, the size of the key and asymmetric ciphers based authentication operations.

Key management issues. Security protocol depends on the security forum. There are two types of security that can be recognized: (1) using a secret key, and (2) use of a certificate authority. If using secret keys, keys to be transferred from one device to another. The key to shipping, we need a set of keys for each pair of devices that need to communicate and be well coordinated. There is also a hardware alternative to this, but these options are expensive and involve a lot of overhead. Digital certificates as an affordable solution to turn coordination is not required as it was in the public key system. Each device requires only a license key and a private key is for management of the installation is stable. However, PKI produces documents as well as having a certain amount of overhead may be unnecessary for smaller systems.

Elliptic Curve Cryptography. Cryptographic Interoperability Strategy (CIS) by the National Security Agency (NSA) approved encryption method of choice for government system started. Of 128 or 256-bit AES encryption is made for. Ephemeral Diffie-Hellman key management model and integrated applications, elliptic curve digital signature algorithm and a secure hash algorithm (SHA) for hashing.

C.    Encryption and key management solutions

In this part we want explain some of advantage and disadvantage of Encryption and key management:

*General design considerations*

•    Encryption should be such that the robust design method and the algorithm is free of defects.

• The number of random bits of entropy can be achieved by seeds deterministic generator (RBG) prior to the distribution or use of the derivative of a function key that comes with the device has been solved.

• The use of cryptographic modules that are used to protect the encryption algorithm. We need time to update this module, the intelligent network can be used for about twenty years and replacing them can be an expensive affair.

• Encryption system failures may occur due to implementation errors, failures, hybrid algorithms, unsafe or improper protocol is secure. These should be considered in the design.

• Of the random number generator is an integral part of the security system, breaking it protocol or encryption algorithm is a compromise.

• Should be replaced by authentication and authorization methods cannot be connected to other systems.

• Availability must always exist as dropping or refusing to re-establish a connection, it may affect the life connection.

• Algorithm and key length should be such that the desired security strength is obtained.

• In order to maintain the security of keying material and data authentication, we need it from unauthorized access or manipulation of devices to protect. Physical security is needed for this purpose.

*Key management system for intelligent networks.*

We are using a certificate that is valid, that the certificate is not expired. If the certificate is issued to a device that is more reliable, or lost or stolen, then the certificate can be revoked. Certificate revocation lists are used for this purpose. A device that uses the information in the certificate, the relying party (RP). Czech RP a list that should be considered in the admission certificate.

The following points need to be checked

• If the trusted CA certificate has been issued;

• If the certificate is valid and not expired;

• The certificate shall be valid CRL,

• Confirmed the issue of certificate policies for which the certificate is being used.

These system can be studied in detail from [13, 14].

The SAGC4ISR systems require appropriate measures and requirements for the vulnerabilities discussed above for achieving confidentiality, integrity and availability of data and maintaining system configuration through elaborate security guidelines and accountability of personnel authorized to access the information sources. Following security services are required to cater the security requirements: Authentication, Access Control, Data Confidentiality, Data Integrity, Non-repudiation, availability that describe all of them in C4ISR security. And also we can use the Security for the Smart Grid information that survey in the past part.

In past part explain autonomic grid in C4ISR and create AGC4ISR architecture, after that explaining the Smart grid and added to AGC4ISR architecture and create the SAGC4ISR architecture. But now we want discussion about the very important problem in SAGC4ISR, it is security.

In this paper we explain some of the security plan that its normal for all the network and architecture but now we want to use them in our problem.

For description this problem, organization all parts connected with security plan that explained in last part.

Now we try to use a security plan in this organization:

First dimension take decisions like echelon from top to bottom in every organization. This decision need to some primary information that after hoarding and spent more time, people can take decision. But now the important problem in this part is security, because it a natural part of C4ISR and should be use the strong security plan in this part. Also propose that use at beginning the cryptography for had the safe packet and for sent to the next layer use the key management.

Second dimension is connected with different systems that say in last part. Also we need a very secure packet in this layer. In last part used the cryptography for all packets, now for connect to another systems can use the key management too.

Third dimension is computing part. In this part smart grid can be get the knowledge and computing them and sent to next and specialist grid. This part Except for saving time makes the calculations to ensure accuracy improve. So this part need to have a secure packet for receive and sent. In last part that used the cryptography for the packet and key management for sending to the next part. So in this part after that computing can use that algorithm again.

Fourth dimension is comported merging and sharing organization that how to survey connect this organization to another organization or even different nationality. This relation is possible vertically with the same organization or horizontally with top or button organization. In this part sent and receive the packet to another organization and use the security plan, dependence to Destination organization. But we propose that use key management, because it is designed for intelligent networks. So percent of the packet or the key is missing it is too few.

Now we can use this security plan for SAGC4ISR. It means that we can match every four parts of AGC4ISR architecture and SAGC4ISR architecture and create a new safe architecture.

IV.     CONCLUSION

To the best of our knowledge, at first we deliberation the Smart grids with advantage and disadvantage of cyber security. In this time, study about some security model.

After that we observation the detail of some cyber security and security plan in Smart Grids and C4ISR.

Finally deliberation the encryption and key management for SAGC4ISR and also result of them is we can use the security plan for SAGC4ISR that discussion in last part.

REFERENCES

[1] D. Dollen," Report to NIST on the Smart Grid Interoperability Standards Roadmap", June 17. 2009.

[2] G. Locke,P. Gallagher "NIST Framework and Roadmap for Smart Grid Interoperability Standards", Office of the National Coordinator for Smart Grid Interoperability , January 2010.

[3] Monica F. Farah-Stapleton et al., "Proposing a C4ISR Architecture Methodology for Homeland Security"

[4] Fraunhofer-InstitutSichereTelekooperation (FhI-SIT).,"Encryption and Digital Sig-natures"

[5] Junaid Ahmed Zubairi. et al.,"Cyber Security Standards,Practices and Industrial Applications: Systems and Methodologies"

[6] G. Iyer and P. Agrawal, "Smart power grids," in Proceedings of the 2010 42nd Southeastern Symposium on System Theory (SSST 2010), pp. 152–155, March 2010.

[7] Z. Vale, H. Morais, P. Faria, H. Khodr, J. Ferreira, and P. Kadar, "Distributed energy resourcesmanagement with cyberphysical SCADA in the context of future smart grids," in Proceedings of the 15th IEEE Mediterranean Electrotechnical Conference (MELECON 2010), pp. 431–436, April 2010.

[8] Mehdi Bahrami, "A Novel Self-Recognition Method for Autonomic Grid Networks Case Study: Advisor Labor Law Software Application." Advances in Information Sciences and Service Sciences 3.5 (2011)..

[9] G. N. Ericsson, "Cyber security and power system communication—essential parts of a smart grid infrastructure," IEEE Transactions on Power Delivery, vol. 25, no. 3, Article ID 5452993, pp. 1501–1507, 2010.

[10] F. Boroomand, A. Fereidunian, M. A. Zamani et al., "Cyber security for smart grid: a human-automation interaction framework," in Proceedings of the IEEE PES Innovative Smart Grid Technologies Conference Europe (ISGT Europe 2010), pp. 1–6, October 2010.

[11] "Cyberspace policy review," http://www.whitehouse.gov/ assets/documents/Cyberspace Policy Review final.pdf.

[12] "Guidelines for smart grid cyber security vol. 1, smart grid cyber security, architecture and high-level requirements," 2010, http://csrc.nist.gov/publications/nistir/ir7628/nistir- 7628 vol1.pdf.

[13] "Advanced encryption standard (AES)," 2001, http://csrc.nist.gov/publications/fips/fips197/fips-197 .pdf.

[14] "Implementation guidance for FIPS PUB 140-2 and the cryptographic module validation," 2010, http://csrc.nist.gov/ groups/STM/cmvp/documents/fips140-2/FIPS1402IG.pdf.

[15] "Computer Networks", Andrew S. Tanenbaum.

[16] Stallings, W. (2003). Cryptography and network security principles and practices (3rd ed.).

[17] "Fault-Tolerant and Scalable Key Management for Smart Grid", Dapeng Wu and Chi Zhou.

[18] Mehdi Bahrami, Ahmad Faraahi, and Amir Masoud Rahmani. "AGC4ISR, New Software Architecture for Autonomic Grid Computing." Intelligent Systems, Modelling and Simulation (ISMS), 2010 International Conference on. IEEE, 2010.

[19] Mehdi Bahrami, Marziyeh Shahrazadfard, and Tooba Kerdkar. "NSSA: A New Enterprise Architecture for Network Setup without Any Network

Infrastructure."Intelligent Systems, Modelling and Simulation (ISMS), 2011 Second International Conference on. IEEE, 2011.

[20] US DoD, "C4ISR Architecture Framework Version 2.0", U.S. DoD Publisher, USA, Dec 1997.

[21] Len Bass, Paul Clements, Rick Kazman, "Software Architecture in Practice", Addison Wesley, Second Edition, 2003.

[22] B. Boehm, "Engineering Context," Proceedings of the First International Workshop on Architectures for Software Systems. Available as CMU-CS-TR-95-151 from the School of Computer Science, Carnegie Mellon University, April 1995.

[23] C4ISR FOR NETWORK-ORIENTED DEFENSE, Ericsson AB 2006, White Paper, October 2006.

[24] "Semantic Web Technologies for a Smart Energy Grid: Requirements and Challenges?",Andreas Wagner, et al.

[25] "INFORMATION AGE ANTHOLOGY: National Security Implications of the Information Age", DAVID S. ALBERTS and DANIEL S. PAPP.